\documentclass[conference]{IEEEtran}
\usepackage{graphicx}%
\usepackage{bm}%
\usepackage{amsfonts}
\usepackage{amssymb}
\usepackage{times}
\usepackage{subfigure}
\usepackage{enumitem}
\usepackage{amsmath} 
\usepackage{latexsym}
\usepackage{hhline}
\usepackage{cite}
\usepackage{caption}
\usepackage{algorithm}
\usepackage{algpseudocode}
\algrenewcommand\algorithmicrequire{\textbf{Input:}}
\algrenewcommand\algorithmicensure{\textbf{Output:}}
\usepackage{multirow} 
\usepackage{xcolor}
\usepackage{epstopdf}
\usepackage{aligned-overset}

\allowdisplaybreaks

\usepackage{geometry}
\usepackage{url}
\newcommand{\TF}{\mathrm{TF}}
\newcommand{\Freq}{\mathrm{F}}
\newcommand{\Delay}{\mathrm{D}}
\newcommand{\DA}{\mathrm{DA}}
\DeclareMathOperator{\asinc}{asinc}
\DeclareMathOperator{\sinc}{sinc}

\begin{document}
\title{Cross-Domain Channel Estimation for MU-MIMO-OFDM \vspace{-2mm}}
\author{\IEEEauthorblockN{
Mingcheng Nie\IEEEauthorrefmark{1},
Shuangyang Li\IEEEauthorrefmark{2}, 
Ruoxi Chong\IEEEauthorrefmark{3}, 
Giuseppe Caire\IEEEauthorrefmark{2}, 
and
Yonghui Li\IEEEauthorrefmark{1} \vspace{1mm}
}
\IEEEauthorblockA{
\IEEEauthorrefmark{1}The University of Sydney, Sydney, Australia\\
\IEEEauthorrefmark{3}Universidad Carlos III de Madrid, Spain\\
\IEEEauthorrefmark{2}Technische Universit{\"a}t Berlin, Berlin, Germany \vspace{-4mm}
}}
\maketitle

\begin{abstract}
Conventional channel estimation for multiple-input multiple-output orthogonal frequency-division multiplexing (MIMO-OFDM) relies on the pilot interpolation without explicitly identifying the
sparse delay--angle (DA) components. To exploit DA sparsity
while maintaining backward compatibility with existing MIMO-OFDM
systems, we propose a cross-domain
channel estimation scheme for multiuser MIMO-OFDM (MU-MIMO-OFDM) systems. The scheme exploits the circular convolution structure in the DA domain to identify candidate DA pairs by correlating delay-domain observations with DA pilot responses. These pairs and the known pilots then define a time–frequency (TF) dictionary for refining the DA coefficients from TF observations through least-squares or sparsity-regularized fitting. The refined coefficients and candidate pairs
reconstruct the aggregate channel across antennas and
subcarriers aligning with the geometric channel model. Numerical
results show approximately $10$~dB and $7$~dB lower
normalized mean-square error than the conventional OFDM estimation schemes for integer and fractional delay and angle indices,
respectively.
\end{abstract}

\section{Introduction}
\label{sec:introduction}

Orthogonal frequency-division multiplexing (OFDM) supports
wideband multiple-input multiple-output (MIMO) transmission by
converting a frequency-selective MIMO channel into a set of
narrowband MIMO subchannels.
Accurate estimates of these subchannels are essential for spatial precoding and detection, which enable parallel transmission and reliable recovery of multiple data streams~\cite{you2016channel}.
Conventional OFDM estimators recover aggregate time--frequency
(TF) channel coefficients at pilot locations and interpolate
those at data locations by exploiting channel correlation
within a coherence region~\cite{coleri2002channel}.

The development of future wireless networks has brought new design opportunities for estimating the MIMO-OFDM channel. For example, sixth-generation (6G) networks are evolving toward upper
mid-band spectrum (frequency range~3, FR3) and extremely
large-scale MIMO arrays~\cite{bazzi2026upper,wang2024tutorial}.
Wider bandwidths and larger array apertures provide finer
delay and angular resolution, respectively. For channels
dominated by a limited number of paths, this finer resolution
enhances the sparsity of the delay--angle (DA) representation,
with most channel energy concentrated around a few
peaks~\cite{you2016channel}.
The frequency--antenna channel can be efficiently transformed
into this sparse DA representation using a pair of fast Fourier
transforms (FFTs). 

In fact, exploiting the advantages of the transform-domains for communications is no longer a new topic.
Noticeably, several waveform designs have emerged to 
exploit channel sparsity in transform-domains. For example, orthogonal time-frequency space
(OTFS) modulation multiplexes symbols in the delay--Doppler
(DD) domain \cite{Li2026OTFSTutorial,Li2026DelayDopplerBook}. Sufficient bandwidth and observation time across
multiple symbols provide fine delay and Doppler resolution,
respectively, yielding an approximately sparse DD representation~\cite{nie2024uplink,nie2026standardizing}.
Embedded-pilot schemes can therefore exploit this DD sparsity to focus channel estimation on a small set of significant components identified by thresholding the received pilot response~\cite{raviteja2019embedded}.
Similarly, affine frequency division multiplexing (AFDM)
exploits the sparse effective channel in the discrete affine
Fourier transform (DAFT) domain for low-complexity
detection~\cite{bemani2023affine}. Orthogonal chirp delay--Doppler
division multiplexing (CDDM) spreads symbols in the DD domain
using the chirp--Zak transform. It exploits DD channel sparsity
to identify path delays and Doppler shifts, then estimates the
corresponding gains through chirp correlation~\cite{bai2026orthogonal}.

Although these waveforms exploit channel sparsity for estimation
and detection, their adoption requires significant transceiver modifications
that can complicate backward compatibility with existing OFDM
systems~\cite{nie2026standardizing}.
Preserving compatibility is particularly important given the adoption of OFDM in fifth-generation New Radio (5G NR)
and its retention in the ongoing 6G radio
study~\cite{nie2026standardizing}.
These practical constraints motivate channel estimation schemes that exploit transform-domain
channel sparsity while preserving the OFDM waveform.
For example, the scheme in~\cite{ma2026ddaofdm} transforms
pilot-based TF channel estimates into the DD domain and uses
maximum likelihood estimation to recover path delays, Doppler
shifts, and gains. These parameters reconstruct the TF channel
matrix for minimum mean-square error equalization that accounts
for intercarrier interference.
In our previous work~\cite{nie2025cdce}, we proposed a two-stage
cross-domain scheme that transforms OFDM pilots and received
samples into the DD domain, where twisted correlation and
thresholding identify candidate DD pairs.
These pairs define a TF dictionary for fading-coefficient
estimation through $\ell_1$-regularized least squares (LS), which suppresses false candidates before channel reconstruction.

In this paper, we extend this cross-domain estimation approach
to multiuser MIMO-OFDM by exploiting the DA sparsity. The proposed
scheme combines coarse DA estimation with TF refinement.
We transform the
known frequency--antenna pilots into the DA domain and the
received samples into the delay domain to exploit DA
sparsity while retaining the transmitted OFDM waveform.
We then identify candidate DA pairs
by correlating the
delay-domain observations with the DA pilot responses and
thresholding the resulting correlation map. These pairs define a TF dictionary for coefficient
estimation using the least absolute shrinkage and selection
operator (LASSO) when the problem is underdetermined, or
LS when the dictionary is well conditioned
and has full column rank. For fractional indices, an
additional LS refit of the nonzero LASSO coefficients reduces
shrinkage bias. The refined coefficients and their associated
DA pairs then reconstruct the aggregate channel
across antennas and subcarriers.
The proposed scheme achieves
approximately $10$~dB and $7$~dB lower normalized mean-square
error (NMSE) than conventional OFDM estimation schemes for integer and fractional path
indices, respectively.

\emph{Notation:} Bold lower- and uppercase letters denote vectors
and matrices. The superscripts $(\cdot)^*$, $(\cdot)^{\rm T}$,
$(\cdot)^{\rm H}$, and $(\cdot)^\dagger$ denote conjugation,
transpose, Hermitian transpose, and the Moore--Penrose
pseudoinverse, respectively.
$[z]_D$ denotes $z$ modulo $D$ in $[0,D)$, and
$\delta_{i,j}$ is the Kronecker delta.

\section{System Model}
\label{sec:model}

\subsection{Downlink Training Model}
We consider a downlink multiuser MIMO-OFDM (MU-MIMO-OFDM) system \cite{Nie2026LearningAidedISAC}.
A base station (BS) with $N_{\rm t}$
transmit antennas broadcasts known pilots to $U$ users, each
equipped with a single receive antenna. The system uses $M$
subcarriers with spacing $\Delta f$ and $N$ sounding OFDM
symbols. The indices $m=0,\ldots,M-1$, $n=0,\ldots,N-1$,
$a=0,\ldots,N_{\rm t}-1$, and $u=1,\ldots,U$ denote subcarriers,
sounding symbols, transmit antennas, and users, respectively.

Let $x_{\TF,a}[n,m]$ denote the known pilot transmitted by antenna
$a$ on subcarrier $m$ of sounding symbol $n$, and let
$y_{\TF,u}[n,m]$ denote the corresponding received sample at user
$u$ after cyclic-prefix (CP) removal and a unitary discrete
Fourier transform (DFT).
The TF coordinates $(n,m)$ identify the sounding symbol and
subcarrier, respectively. We assume independent noise samples
$w_{\TF,u}[n,m]\sim\mathcal{CN}(0,N_{0})$ and neglect
interference during sounding.

We assume a time-invariant channel over the $N$ sounding
symbols. The frequency response $h_{\Freq,u}[a,m]$ from
antenna $a$ to user $u$ is therefore independent of $n$,
but may vary across subcarriers. For each subcarrier $m$,
we collect the channel coefficients across antennas into
$\mathbf h_{\Freq,u}[m]=[h_{\Freq,u}[0,m],\ldots,
h_{\Freq,u}[N_{\rm t}-1,m]]^{\rm T}$ and the observations
across sounding symbols into
$\mathbf y_{\TF,u}[m]=[y_{\TF,u}[0,m],\ldots,
y_{\TF,u}[N-1,m]]^{\rm T}$.
The corresponding pilot matrix
$\mathbf X_{\TF}[m]\in\mathbb C^{N\times N_{\rm t}}$
has entries $[\mathbf X_{\TF}[m]]_{n,a}=x_{\TF,a}[n,m]$.
We collect the noise samples over the $N$ sounding symbols
into the vector $\mathbf w_{\TF,u}[m]
=[w_{\TF,u}[0,m],\ldots,w_{\TF,u}[N-1,m]]^{\rm T}$.
The downlink training relation is then given by \cite{song2025csit}
\begin{equation}
 \mathbf y_{\TF,u}[m]
 =\mathbf X_{\TF}[m]\mathbf h_{\Freq,u}[m]+\mathbf w_{\TF,u}[m].
 \label{eq:subcarrier-io}
\end{equation}
We next collect the frequency-domain channel coefficients
over all $M$ subcarriers into
$\mathbf h_{\Freq,u}=[\mathbf h_{\Freq,u}[0]^{\rm T},\ldots,
\mathbf h_{\Freq,u}[M-1]^{\rm T}]^{\rm T}\in\mathbb C^{MN_{\rm t}}$.
Similarly, we stack the observations as
$\mathbf y_{\TF,u}=[\mathbf y_{\TF,u}[0]^{\rm T},\ldots,
\mathbf y_{\TF,u}[M-1]^{\rm T}]^{\rm T}\in\mathbb C^{MN}$
and define $\mathbf w_{\TF,u}$ in the same way. The block-diagonal
pilot matrix is $\mathbf X_{\TF}=\operatorname{blkdiag}
\big(\mathbf X_{\TF}[0],\ldots,\mathbf X_{\TF}[M-1]\big)
\in\mathbb C^{MN\times MN_{\rm t}}$.
Stacking \eqref{eq:subcarrier-io} over the $M$ subcarriers gives
\begin{equation}
 \mathbf y_{\TF,u}=\mathbf X_{\TF}\mathbf h_{\Freq,u}+\mathbf w_{\TF,u},
 \label{eq:pilot-channel-io}
\end{equation}
where $\mathbf X_{\TF}$ denotes the pilots broadcast by the BS to all users, and the stacked noise satisfies
$\mathbf w_{\TF,u}\sim\mathcal{CN}(\mathbf0,N_{0}\mathbf I_{MN})$.

\subsection{Geometric Channel Model}
We consider a uniform linear array (ULA) with
half-wavelength element spacing at the BS and assume far-field
propagation and a narrowband array response. For the $u$-th user, the channel vector is \cite{song2025csit}
\begin{equation}
 \mathbf h_{\Freq,u}=\operatorname{vec}\!\left(
 \sum_{p=1}^{P_u}g_{u,p}
 \mathbf a(\theta_{u,p})\mathbf b^{\rm T}(\tau_{u,p})
 \right),
 \label{eq:geometric-vector}
\end{equation}
where $g_{u,p}$, $\tau_{u,p}$, and $\theta_{u,p}$ are the
complex fading coefficient, delay, and angle of departure
(AoD) of path $p$. The vectors
$\mathbf a(\theta)\in\mathbb C^{N_{\rm t}}$ and
$\mathbf b(\tau)\in\mathbb C^M$ are defined by
\begin{equation}
 \begin{aligned}
 \mathbf a(\theta)
 &=[1,e^{j\pi\sin\theta},\ldots,e^{j\pi(N_{\rm t}-1)\sin\theta}]^{\rm T},\\
 \mathbf b(\tau)
 &=[1,e^{-j2\pi\Delta f\tau},\ldots,e^{-j2\pi(M-1)\Delta f\tau}]^{\rm T}.
 \end{aligned}
 \label{eq:steering-delay}
\end{equation}
Here, $\mathbf a(\theta)$ and $\mathbf b(\tau)$ describe the
spatial and frequency responses, respectively. Column-wise
vectorization in \eqref{eq:geometric-vector} groups the
antenna coefficients within each subcarrier. The corresponding
scalar channel coefficient is
\begin{equation}
 h_{\Freq,u}[a,m]=\sum_{p=1}^{P_u}g_{u,p}
 e^{j\pi a\sin\theta_{u,p}}
 e^{-j2\pi m\Delta f\tau_{u,p}}.
 \label{eq:channel}
\end{equation}

The path parameters $g_{u,p}$, $\tau_{u,p}$, and $\theta_{u,p}$
remain constant throughout the $N$ sounding symbols and $M$ subcarriers. We
assume perfect timing and carrier synchronization and
rectangular OFDM with
$0\leq\tau_{u,p}\leq\tau_{\max}\leq T_{\rm CP}<1/\Delta f$,
so the CP accommodates the delay spread and the
frequency-domain model is free of intersymbol and
intercarrier interference \cite{tse2005fundamentals}.
The AoDs lie in a known sector
$|\theta_{u,p}|\leq\theta_{\max}<\pi/2$.

\subsection{Symbol-Wise Input--Output Relation}
We now express each received sample in terms of the path
parameters. Taking row $n$ of \eqref{eq:subcarrier-io} and
substituting \eqref{eq:channel} gives
\begin{equation}
 \begin{split}
 y_{\TF,u}[n,m]
 &=\sum_{a=0}^{N_{\rm t}-1}x_{\TF,a}[n,m]h_{\Freq,u}[a,m]+w_{\TF,u}[n,m]\\
 &=\sum_{p=1}^{P_u}g_{u,p}e^{-j2\pi m\Delta f\tau_{u,p}}\\
 &\quad{}\times\sum_{a=0}^{N_{\rm t}-1}
 x_{\TF,a}[n,m]e^{j\pi a\sin\theta_{u,p}}+w_{\TF,u}[n,m].
 \end{split}
 \label{eq:raw}
\end{equation}
In \eqref{eq:raw}, each path introduces linear phase
variations across subcarriers and BS antennas, determined
by its delay and AoD, respectively. This motivates us to express
the pilots in the DA domain and the observations in the
delay domain to characterize how each path shifts the DA pilot coefficients.

\section{Cross Domain Channel Estimation}
\label{sec:correlation}
Conventional OFDM schemes employ linear or Wiener
interpolation to recover channel coefficients on non-pilot
subcarriers from pilot-based channel estimates
\cite{coleri2002channel,hoeher1997wiener}. We instead estimate
the aggregate channel through an equivalent DA representation
determined by its path gains, delays, and AoDs. These path
parameters remain invariant over the modeled bandwidth and
sounding interval under the assumptions of Section~\ref{sec:model}.
This approach offers two advantages. First, it exploits DA sparsity to focus estimation on a small set of significant components that jointly describe the channel across antennas and subcarriers. Second, it reconstructs the channel across antennas
and subcarriers based on selected equivalent paths,
providing a physical basis for interpolation through the
geometric channel model. We then derive the general input--output relation in the DA domain and explain the cross-domain framework.


\subsection{Input--Output Relation in the Delay--Angle Domain}
To expose the delay and AoD dependence in \eqref{eq:raw},
we apply the symplectic finite Fourier transform
(SFFT)~\cite{nie2026standardizing} across the subcarrier
and transmit-antenna dimensions of the known pilots.
At each single-antenna user, an inverse DFT across
subcarriers yields delay-domain observations. For sounding
symbol $n$, the resulting DA pilot samples and
delay-domain observations are given by
\begin{align}
 X_{\DA,n}[l,q]
 &=\frac{1}{\sqrt{MN_{\rm t}}}
 \sum_{a=0}^{N_{\rm t}-1}\sum_{m=0}^{M-1}x_{\TF,a}[n,m]\nonumber\\
 &\quad{}\times e^{j2\pi ml/M}e^{-j2\pi aq/N_{\rm t}},
 \label{eq:da-x}\\
 y_{\Delay,u,n}[l]
 &=\frac{1}{\sqrt M}\sum_{m=0}^{M-1}
 y_{\TF,u}[n,m]e^{j2\pi ml/M},
 \label{eq:delay-y}
\end{align}
where $l=0,\ldots,M-1$ and $q=0,\ldots,N_{\rm t}-1$
index the delay and angular bins, respectively. Moreover,
we define each physical path's normalized delay and angle
indices as
\begin{equation}
 l_{u,p}=M\Delta f\tau_{u,p},\qquad
 q_{u,p}=\frac{N_{\rm t}}{2}\sin\theta_{u,p}.
 \label{eq:path-coordinates}
\end{equation}

We neglect noise during the intermediate derivation and
restore it in the final relation. Substituting
the scalar input--output relation \eqref{eq:raw} into the
delay transform \eqref{eq:delay-y} gives
\begin{equation*}
 \begin{split}
 y_{\Delay,u,n}[l]
 &=\frac{1}{\sqrt M}\sum_{m=0}^{M-1}
 y_{\TF,u}[n,m]e^{j2\pi ml/M}\\
 &=\frac{1}{\sqrt M}\sum_{m=0}^{M-1}
 \sum_{a=0}^{N_{\rm t}-1}x_{\TF,a}[n,m]h_{\Freq,u}[a,m]e^{j2\pi ml/M}.
 \end{split}
\end{equation*}
Using \eqref{eq:channel} together with
$\Delta f\tau_{u,p}=l_{u,p}/M$ and
$\sin\theta_{u,p}=2q_{u,p}/N_{\rm t}$ from
\eqref{eq:path-coordinates}, we obtain
\begin{equation}
 \begin{split}
 y_{\Delay,u,n}[l]
 &=\frac{1}{\sqrt M}\sum_{m=0}^{M-1}
 \sum_{a=0}^{N_{\rm t}-1}\sum_{p=1}^{P_u}
 g_{u,p}x_{\TF,a}[n,m]\\
 &\quad{}\times e^{-j2\pi m l_{u,p}/M}
 e^{j2\pi a q_{u,p}/N_{\rm t}}e^{j2\pi ml/M}.
 \end{split}
 \label{eq:da-substitution}
\end{equation}
To express \eqref{eq:da-substitution} in terms of the known
DA pilot samples, we apply the corresponding inverse SFFT
(ISFFT):
\begin{equation}
 \begin{split}
 x_{\TF,a}[n,m]
 &=\frac{1}{\sqrt{MN_{\rm t}}}
 \sum_{l'=0}^{M-1}\sum_{q'=0}^{N_{\rm t}-1}
 X_{\DA,n}[l',q']\\
 &\quad{}\times e^{-j2\pi ml'/M}e^{j2\pi aq'/N_{\rm t}}.
 \end{split}
 \label{eq:da-x-inverse}
\end{equation}
Substituting \eqref{eq:da-x-inverse} into
\eqref{eq:da-substitution} yields
\begin{equation}
 \begin{split}
 &y_{\Delay,u,n}[l]
 =\frac{1}{M\sqrt{N_{\rm t}}}
 \sum_{m=0}^{M-1}\sum_{a=0}^{N_{\rm t}-1}\sum_{p=1}^{P_u}
 g_{u,p}\\
 &\quad{}\times\sum_{l'=0}^{M-1}\sum_{q'=0}^{N_{\rm t}-1}
 X_{\DA,n}[l',q']e^{-j2\pi ml'/M}e^{j2\pi aq'/N_{\rm t}}\\
 &\quad{}\times e^{-j2\pi m l_{u,p}/M}
 e^{j2\pi a q_{u,p}/N_{\rm t}}e^{j2\pi ml/M}.
 \end{split}
 \label{eq:da-inverse-substitution}
\end{equation}
After some manipulation, we restore the noise
term and obtain
\begin{equation}
 \begin{split}
 y_{\Delay,u,n}[l]
 &=\frac{1}{M\sqrt{N_{\rm t}}}
 \sum_{p=1}^{P_u}g_{u,p}
 \sum_{l'=0}^{M-1}\sum_{q'=0}^{N_{\rm t}-1}
 X_{\DA,n}[l',q']\\
 &\quad{}\times
 \left(\sum_{m=0}^{M-1}e^{j2\pi m(l-l_{u,p}-l')/M}\right)\\
 &\quad{}\times
 \left(\sum_{a=0}^{N_{\rm t}-1}
 e^{j2\pi a(q'+q_{u,p})/N_{\rm t}}\right)
 +w_{\Delay,u,n}[l].
 \end{split}
 \label{eq:da-io}
\end{equation}
Equation~\eqref{eq:da-io} expresses each delay-domain
observation as a weighted sum of DA pilot samples for
arbitrary real-valued delay and angle indices. For each path,
the weights are determined by its gain and the two finite
Fourier sums associated with the delay and angular
offsets. Since the path parameters remain unchanged over
the $N$ sounding symbols, the corresponding observations
can be used jointly for channel estimation. We next evaluate
the finite Fourier sums for integer and fractional path
indices.

\subsubsection{Integer path indices}
We first consider paths aligned with the DA grid, i.e.,
$l_{u,p},q_{u,p}\in\mathbb Z$. In this case, both Fourier
sums in \eqref{eq:da-io} can be evaluated by finite Fourier
orthogonality. Specifically, for any integer $d$ and positive
integer $K$, we have
\begin{equation*}
 \sum_{k=0}^{K-1}e^{j2\pi kd/K}
 =\begin{cases}
 K,&d\in K\mathbb Z,\\
 0,&\text{otherwise}.
 \end{cases}
\end{equation*}
Indeed, when $d\in K\mathbb Z$, every term in the sum equals one.
Otherwise, the formula for a finite geometric sum gives
$(1-e^{j2\pi d})/(1-e^{j2\pi d/K})=0$ because its numerator
vanishes and its denominator is nonzero. Applying this
identity directly to \eqref{eq:da-io} yields
\begin{equation*}
 \begin{aligned}
 \sum_{m=0}^{M-1}e^{j2\pi m(l-l_{u,p}-l')/M}
 &=M\delta_{l',[l-l_{u,p}]_M},\\
 \sum_{a=0}^{N_{\rm t}-1}e^{j2\pi a(q'+q_{u,p})/N_{\rm t}}
 &=N_{\rm t}\delta_{q',[-q_{u,p}]_{N_{\rm t}}}.
 \end{aligned}
\end{equation*}
The first sum is nonzero only when
$l-l_{u,p}-l'\in M\mathbb Z$. Since $0\leq l'<M$, the
unique matching index is $l'=[l-l_{u,p}]_M$. The same
argument selects $q'=[-q_{u,p}]_{N_{\rm t}}$ in the second
sum. The two selectors therefore reduce \eqref{eq:da-io} to
\begin{equation}
 \begin{split}
 &y_{\Delay,u,n}[l]
 =\sqrt{N_{\rm t}}\sum_{p=1}^{P_u}g_{u,p}\\
 &\quad{}\times X_{\DA,n}\big[[l-l_{u,p}]_M,
 [-q_{u,p}]_{N_{\rm t}}\big]
 +w_{\Delay,u,n}[l].
 \end{split}
 \label{eq:da-io-integer}
\end{equation}
Equation~\eqref{eq:da-io-integer} shows that the path delay
and AoD appear as shifts in the delay and angular indices
of the DA pilots, respectively, while $\sqrt{N_{\rm t}}g_{u,p}$
scales each path contribution. The modulo operations express
these shifted indices within the finite DA grid.

\subsubsection{Fractional path indices}
Physical delays and AoDs can fall between DA grid points.
We therefore allow noninteger delay and angle indices and
decompose them into their integer and fractional parts:
\begin{equation*}
 l_{u,p}=L_{u,p}+\epsilon_{u,p},\qquad
 q_{u,p}=Q_{u,p}+\eta_{u,p},
\end{equation*}
where $L_{u,p},Q_{u,p}\in\mathbb Z$ and
$\epsilon_{u,p},\eta_{u,p}\in[-1/2,1/2)$ are the fractional components. A nonzero
$\epsilon_{u,p}$ or $\eta_{u,p}$ prevents the corresponding
Fourier sum in \eqref{eq:da-io} from selecting a single
pilot index. We next evaluate the two Fourier sums in
\eqref{eq:da-io} for fractional path indices.

For a positive integer $K$ and a real
$t\notin K\mathbb Z$, the finite geometric sum gives
\begin{equation*}
 \sum_{k=0}^{K-1}e^{j2\pi kt/K}
 =\frac{1-e^{j2\pi t}}{1-e^{j2\pi t/K}}
 =e^{j\pi(K-1)t/K}
 \frac{\sin(\pi t)}{\sin(\pi t/K)}.
\end{equation*} 
Define the normalized aliased sinc function by
\begin{equation*}
 \asinc_K(x)=\frac{\sin(\pi Kx)}{K\sin(\pi x)},
 \qquad x\notin\mathbb Z.
\end{equation*}
For $x\in\mathbb Z$, we define $\asinc_K(x)=(-1)^{(K-1)x}$
by continuity. In particular, $\asinc_K(0)=1$.
Therefore, for every real $t$,
\begin{equation*}
 \sum_{k=0}^{K-1}e^{j2\pi kt/K}
 =K e^{j\pi(K-1)t/K}\asinc_K\!\left(\frac{t}{K}\right).
\end{equation*}

Applying the above identity with $K=M$ and $t=l-l_{u,p}-l'$
to the subcarrier sum, and with $K=N_{\rm t}$ and
$t=q'+q_{u,p}$ to the antenna sum, we rewrite
\eqref{eq:da-io} as
\begin{equation}
 \begin{split}
 &y_{\Delay,u,n}[l]
 =\sqrt{N_{\rm t}}\sum_{p=1}^{P_u}g_{u,p}
 \sum_{l'=0}^{M-1}\sum_{q'=0}^{N_{\rm t}-1}
 X_{\DA,n}[l',q']\\
 &\quad{}\times e^{j\pi(M-1)(l-l_{u,p}-l')/M}
 \asinc_M\!\left(\frac{l-l_{u,p}-l'}{M}\right)\\
 &\quad{}\times e^{j\pi(N_{\rm t}-1)(q'+q_{u,p})/N_{\rm t}}
 \asinc_{N_{\rm t}}\!\left(\frac{q'+q_{u,p}}{N_{\rm t}}\right)\\
 &\quad{}+w_{\Delay,u,n}[l].
 \end{split}
 \label{eq:da-io-asinc}
\end{equation}
The two asinc factors in \eqref{eq:da-io-asinc} characterize
delay and angle leakage. A noninteger delay or angle index
produces nonzero Fourier weights at multiple bins in the
corresponding dimension. The delay main-lobe width is
proportional to $1/(M\Delta f)$.
At a fixed AoD, the angle main-lobe width
is asymptotically proportional to $1/N_{\rm t}$
\cite{tse2005fundamentals}. Thus, as
$M$ and $N_{\rm t}$ approach infinity with fixed subcarrier
and antenna spacings, the main lobes narrow, providing
increasingly fine delay and angle resolution, respectively.

Near its central peak, the asinc function can be approximated
by the sinc function, i.e.,
$\asinc_K(\xi/K)\approx\sinc(\xi)$ for $|\xi|\ll K$,
where $\xi$ is
the offset in grid bins and
$\sinc(\xi)=\sin(\pi\xi)/(\pi\xi)$, with $\sinc(0)=1$.
Here, $K=M$ for delay and $K=N_{\rm t}$ for angle.
The approximation is valid locally for finite
$M$ and $N_{\rm t}$ and becomes exact as $K\to\infty$
for fixed $\xi$.
Periodicity gives the same magnitude profile around the
other peaks. The sinc function has a main lobe and decaying sidelobes.
Consequently, noninteger delay or angle indices can cause
significant leakage into neighboring bins even as $M$ and
$N_{\rm t}$ increase. This motivates detecting peaks above
a threshold on the DA grid and retaining coefficients in
neighboring bins to capture the main lobe and significant
sidelobes. The number of neighboring bins retained should
be sufficient to capture significant leakage while limiting
the inclusion of noise bins.

\subsection{Coarse Estimation in the Delay--Angle Domain}
For coarse estimation, we exploit the structure of
\eqref{eq:da-io-integer}: each path with integer delay and
angle indices contributes a scaled and shifted version of
the DA pilots, with the scaling determined by its gain and
the index shifts determined by its delay and AoD. Since
the shifted pilot response is known for each candidate
DA pair, we correlate it with the received
delay-domain samples to identify candidate pairs for
subsequent refinement. Specifically, for
$l=0,\ldots,M-1$ and $q=0,\ldots,N_{\rm t}-1$, we define
the per-user DA correlation map as
\begin{equation}
 \begin{split}
 V_{\DA,u}[l,q]
 &=\frac{\sqrt{N_{\rm t}}}{N}\sum_{n=0}^{N-1}\sum_{r=0}^{M-1}
 y_{\Delay,u,n}^{*}[r]\\
 &\quad{}\times X_{\DA,n}\big[[r-l]_M,[-q]_{N_{\rm t}}\big].
 \end{split}
 \label{eq:corr}
\end{equation}
For each sounding symbol, \eqref{eq:corr} correlates the
received delay-domain samples with the corresponding
DA pilot response. The factor $1/N$ averages these
correlations over the $N$ symbols to reduce the effect
of independent receiver noise.

We next threshold $|V_{\DA,u}[l,q]|$ to identify candidate
DA pairs. The correlation
in \eqref{eq:corr} also applies to fractional physical path
indices, for which \eqref{eq:da-io-asinc} expresses each path
contribution through multiple DA grid responses. We therefore
retain significant neighboring bins around the detected
peaks to capture fractional leakage. Specifically, the pilot-response energy at bin $(l,q)$ is
\begin{equation*}
 E_{\DA}[l,q]
 =N_{\rm t}\sum_{n=0}^{N-1}\sum_{r=0}^{M-1}\left|
 X_{\DA,n}\big[[r-l]_M,[-q]_{N_{\rm t}}\big]
 \right|^2.
\end{equation*}
With the averaging in \eqref{eq:corr}, the noise contribution
has variance
\begin{equation*}
 N_{V}[l,q]=\frac{N_{0} E_{\DA}[l,q]}{N^2}.
\end{equation*}
Following \cite{raviteja2019embedded}, we retain bin $(l,q)$ if
\begin{equation*}
 |V_{\DA,u}[l,q]|>3\sqrt{N_{V}[l,q]}.
\end{equation*}
Let $(\widehat l_{u,p},\widehat q_{u,p})$,
$p=1,\ldots,\widehat P_u$, denote the retained DA
pairs. When the physical paths and their fractional leakage
are detected, $\widehat P_u$ is expected to exceed the physical
path count $P_u$, since one path can contribute several
retained bins.

\subsection{Time--Frequency Refinement and Reconstruction}
Following coarse DA estimation, we refine the channel estimate
in the TF domain by selecting a subset of the candidate
DA pairs and estimating their fading coefficients.
When path indices are fractional,
the retained integer bins may not accurately represent the
corresponding TF responses. To reduce this grid mismatch,
we form a finer local grid around each retained DA
pair. A finer grid can reduce mismatch but increases dictionary
size and column correlations. Let $(\bar l_{u,p},\bar q_{u,p})$, $p=1,\ldots,K_u$, denote
the candidate DA pairs, which coincide with the
$\widehat P_u$ pairs retained by coarse DA estimation when no grid refinement
is applied. Let $\mathbf h_u\in\mathbb C^{K_u}$ collect the DA coefficients
associated with the candidate pairs.
Based on \eqref{eq:raw},
we define the dictionary
$\mathbf D_{\TF,u}\in\mathbb C^{MN\times K_u}$
that maps these DA coefficients to the TF observations as
\begin{equation*}
 [\mathbf D_{\TF,u}]_{i,p}
 =e^{-j2\pi m\bar l_{u,p}/M}\sum_{a=0}^{N_{\rm t}-1}
 x_{\TF,a}[n,m]e^{j2\pi a\bar q_{u,p}/N_{\rm t}}.
\end{equation*}
Here, row $i=mN+n+1$ corresponds to subcarrier $m$ and
sounding symbol $n$, while column $p$ contains the TF pilot response for the candidate pair
$(\bar l_{u,p},\bar q_{u,p})$ with a unit DA coefficient.
Given a candidate DA coefficient estimate
$\widehat{\mathbf h}_u\in\mathbb C^{K_u}$, we formulate the following
LS problem:
\begin{equation}
 \underset{\widehat{\mathbf h}_u}{\operatorname{min}}\;
 \frac12\left\|\mathbf y_{\TF,u}
 -\mathbf D_{\TF,u}\widehat{\mathbf h}_u\right\|_2^2.
 \label{eq:tf-ls}
\end{equation}
A unique LS estimate requires $\mathbf D_{\TF,u}$ to have
full column rank, which requires $MN\geq K_u$. When $MN<K_u$,
the problem is underdetermined and the coefficient estimates
are nonunique. In this case, we estimate the coefficients and select
a subset of the candidate DA pairs by solving the
LASSO problem
\begin{equation}
 \widetilde{\mathbf h}_u
 =\underset{\mathbf h_u}{\operatorname{arg\,min}}\Bigl\{
 \frac12\left\|\mathbf y_{\TF,u}
 -\mathbf D_{\TF,u}\mathbf h_u\right\|_2^2
 +\lambda\|\mathbf h_u\|_1\Bigr\},
 \label{eq:tf-coefficient-estimation}
\end{equation}
where $\lambda>0$ controls the sparsity penalty.
A LASSO solver can be found
in~\cite{nie2025cdce}. Alternatively, when $MN\geq K_u$, we solve \eqref{eq:tf-ls}
directly, provided that $\mathbf D_{\TF,u}$ has full column
rank and is well conditioned. The unique LS estimate is
given by the pseudoinverse~\cite{nie2025cdce}:
\begin{equation*}
 \widehat{\mathbf h}_u
 =\mathbf D_{\TF,u}^{\dagger}\mathbf y_{\TF,u}
 =\left(\mathbf D_{\TF,u}^{\rm H}\mathbf D_{\TF,u}\right)^{-1}
 \mathbf D_{\TF,u}^{\rm H}\mathbf y_{\TF,u}.
\end{equation*}
For fractional path indices, we re-estimate the nonzero
LASSO coefficients by LS to reduce shrinkage, keeping their
DA pairs fixed and the remaining coefficients zero. Finally, we obtain the equivalent delays and AoDs from the
$K_u$ candidate pairs using \eqref{eq:path-coordinates}.
The aggregate channel is then reconstructed from their
estimated coefficients as
\begin{equation}
 \widehat{\mathbf h}_{\Freq,u}
 =\operatorname{vec}\!\left(
 \sum_{p=1}^{K_u}\widehat g_{u,p}
 \mathbf a(\widehat\theta_{u,p})
 \mathbf b^{\rm T}(\widehat\tau_{u,p})
 \right).
 \label{eq:channel-reconstruction-equivalent}
\end{equation}



\section{Numerical Results}
\label{sec:numerical-results}
In this section, we compare the NMSE of the proposed scheme
with that of conventional OFDM estimators for integer and
fractional delay and angle indices.

\subsection{Simulation Setup}
\begin{figure}[t]
    \centering
    \includegraphics[scale=0.5]{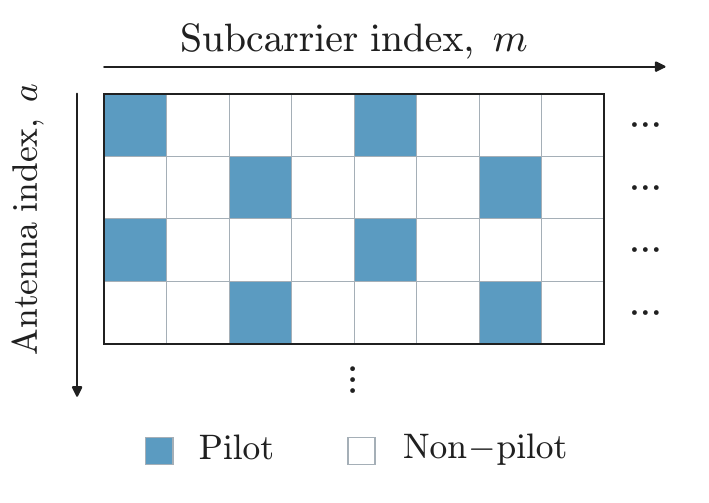}
    \caption{Pilot placement in the frequency--antenna domain. Antenna $a$ transmits in sounding symbol $n=a$.}
    \label{fig:pilot_placement}
\end{figure}

We consider a BS with $N_{\rm t}=8$ half-wavelength-spaced
ULA elements serving $U=4$ single-antenna users over $M=64$
subcarriers. Each sounding interval contains $N=8$ OFDM
symbols with a CP of four samples. Each user's channel
comprises $P_u=3$ physical paths with independent gains
$g_{u,p}\sim\mathcal{CN}(0,1/P_u)$, giving unit average power
per channel coefficient. The delay indices lie in $[0,4]$,
and the AoDs lie in $[-60^\circ,60^\circ]$.
The BS transmits unit-energy quadrature phase-shift keying
(QPSK) pilots using time-division sounding, with antenna $a$
active in symbol $n=a$, as illustrated in
Fig.~\ref{fig:pilot_placement}. The pilot signal-to-noise
ratio (SNR) is $1/N_{0}$, and coarse estimation uses the
threshold $3\sqrt{N_{V}}$. TF refinement uses 128 pilot observations and, for fractional
indices, a local candidate grid with a spacing of $0.1$ in
both the delay and angle indices. We use direct LS
when the candidate dictionary has full column rank and is
well conditioned; otherwise, we solve the LASSO problem in
\eqref{eq:tf-coefficient-estimation}. For fractional indices,
we refit the selected coefficients by LS after LASSO converges,
provided that the selected dictionary satisfies the same rank
and conditioning requirements. We measure channel
reconstruction error over all antennas and subcarriers using
\begin{equation}
 \operatorname{NMSE}
 =\mathbb E\!\left[
 \frac{\|\widehat{\mathbf h}_{\Freq,u}-\mathbf h_{\Freq,u}\|_2^2}
 {\|\mathbf h_{\Freq,u}\|_2^2}\right].
 \label{eq:nmse}
\end{equation}

We consider two benchmarks. The \emph{LS + linear}
benchmark estimates each antenna's channel at its pilot
subcarriers using LS and linearly interpolates these
estimates across frequency.
The linear minimum mean-square error (LMMSE) benchmark,
labeled \emph{LMMSE + Wiener}, estimates the channel at pilot positions using LMMSE estimation and then applies Wiener interpolation to non-pilot positions. Both operations use a channel covariance matrix computed analytically from the assumed distributions of the path gains, delays, and angles.

\subsection{NMSE Performance}
\begin{figure}[t]
    \centering
    \includegraphics[scale=0.4]{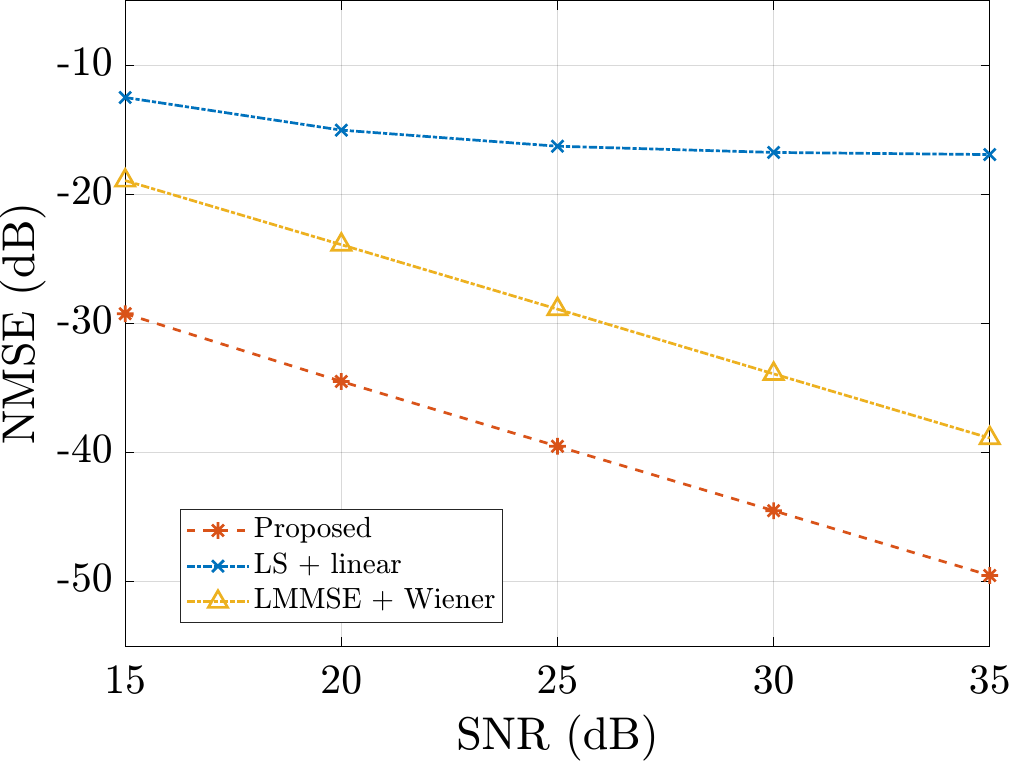}
    \caption{NMSE versus SNR for integer delay and angle indices.}
    \label{fig:nmse_integer}
\end{figure}

Fig.~\ref{fig:nmse_integer} compares the NMSE for integer
delay and angle indices. The proposed scheme's NMSE decreases
with SNR and remains approximately $10$~dB below LMMSE + Wiener
across the evaluated SNR conditions. LS + linear achieves only marginal NMSE
reductions at high SNR because different path delays produce a channel
response that generally does not vary linearly between
pilot subcarriers. Increasing the SNR therefore improves
the pilot estimates but does not eliminate the error
introduced by linear interpolation. LMMSE + Wiener filtering
improves reconstruction by exploiting frequency--antenna
correlation, but does not explicitly identify the sparse DA
components in each channel realization. In contrast, the
proposed scheme exploits DA sparsity by selecting candidate
components through correlation and thresholding. It then
refines their coefficients using TF observations and
reconstructs the channel according to the geometric model.

\begin{figure}[t]
    \centering
    \includegraphics[scale=0.4]{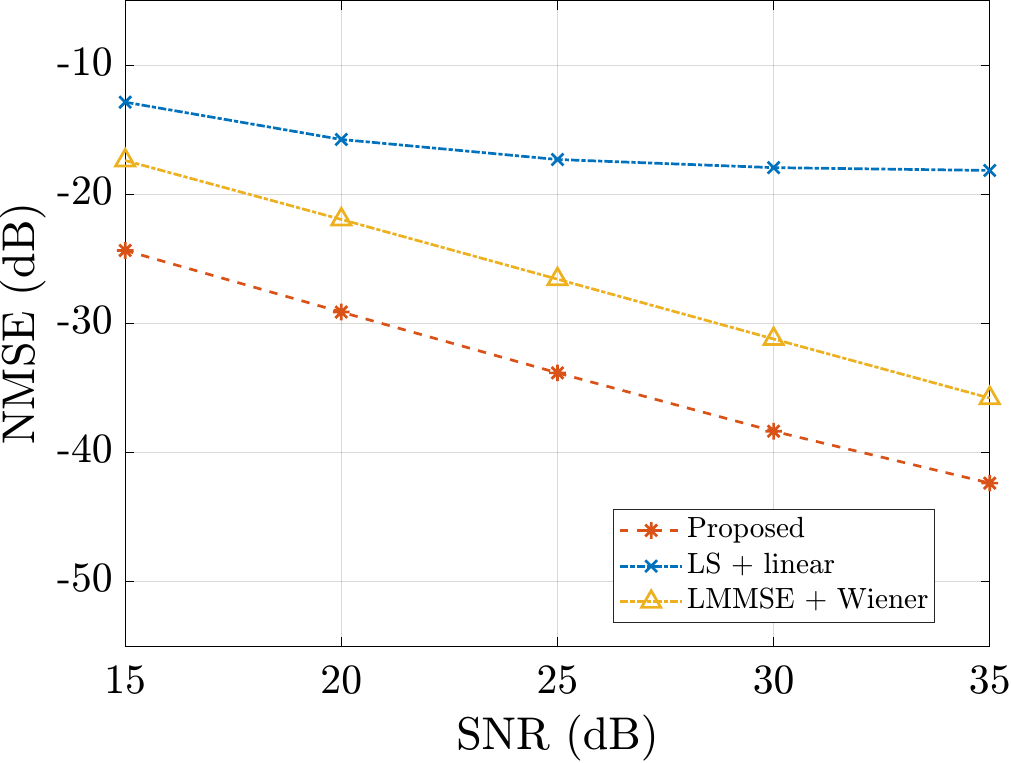}
    \caption{NMSE versus SNR for fractional delay and angle indices.}
    \label{fig:nmse_fractional}
\end{figure}

Fig.~\ref{fig:nmse_fractional} presents the results for
fractional delay and angle indices. The proposed scheme
achieves approximately $7$~dB lower NMSE than LMMSE + Wiener,
while LS + linear again achieve only
marginal NMSE reductions at high SNR. Compared with the integer case, the NMSE is higher and decreases less rapidly as the SNR increases. This behavior is consistent with fractional
leakage and residual mismatch between the physical path
indices and the finer local grid. Nevertheless, the proposed
scheme maintains lower NMSE than both benchmarks
under all evaluated SNR conditions.

\section{Conclusion}
\label{sec:conclusion}
We developed a cross-domain channel estimation scheme for
MU-MIMO-OFDM that combines coarse DA estimation with TF
refinement. It exploits DA channel sparsity while maintaining
backward compatibility with existing OFDM systems.
We first identify candidate DA pairs through
correlation and thresholding, then construct a TF dictionary
from these pairs and the known pilots. Using this dictionary,
we refine the DA coefficients by fitting the corresponding
pilot responses to the TF observations. The aggregate channel
is then reconstructed across antennas and subcarriers from
the refined coefficients and candidate pairs.
Numerical results show approximately $10$~dB and $7$~dB lower
NMSE than the best of the conventional OFDM estimation schemes
considered for integer and fractional path indices, respectively.

\vspace{-2mm}

\bibliographystyle{IEEEtran}
\bibliography{ref}
\end{document}